\documentclass[a4paper,twocolumn]{aa}
\usepackage{graphicx}
\usepackage{natbib}
\begin{document}


\def\be{\begin{equation}}
\def\ee{\end{equation}}
\def\bea{\begin{eqnarray}}
\def\eea{\end{eqnarray}}
\def\c{\cite}

\def\et{ {\it et al.}}
\def\lan{ \langle}
\def\ran{ \rangle}
\def\ov{ \over}
\def\ep{ \epsilon}

\def\mdot{\ifmmode \dot M \else $\dot M$\fi}    
\def\mxd{\ifmmode \dot {M}_{x} \else $\dot {M}_{x}$\fi}
\def\med{\ifmmode \dot {M}_{Edd} \else $\dot {M}_{Edd}$\fi}
\def\bff{\ifmmode B_{f} \else $B_{f}$\fi}

\def\iaucirc{{\it IAU\,Circ. No.}}       
\def\apj{\ifmmode ApJ \else ApJ \fi}    
\def\apjl{\ifmmode  ApJ \else ApJ \fi}    %
\def\aap{\ifmmode A\&A \else A\&A\fi}    %
\def\mnras{\ifmmode MNRAS \else MNRAS \fi}    %
\def\nat{\ifmmode Nature \else Nature \fi}
\def\prl{\ifmmode Phys. Rev. Lett. \else Phys. Rev. Lett.\fi}
\def\prd{\ifmmode Phys. Rev. D. \else Phys. Rev. D.\fi}

\def\la{ \langle}

\def\ms{\ifmmode {\rm M_{\odot}} \else ${\rm M_{\odot}}$\fi}    
\def\na{\ifmmode \nu_{\rm A} \else $\nu_{\rm A}$\fi}    
\def\nk{\ifmmode \nu_{\rm k} \else $\nu_{\rm k}$\fi}    
\def\ns{\ifmmode \nu_{{\rm s}} \else $\nu_{{\rm s}}$\fi}
\def\no{\ifmmode \nu_{1} \else $\nu_{1}$\fi}    
\def\nt{\ifmmode \nu_{2} \else $\nu_{2}$\fi}    
\def\ntk{\ifmmode \nu_{\rm 2k} \else $\nu_{\rm 2k}$\fi}    
\def\dnmax{\ifmmode \Delta \nu_{\rm max} \else $\Delta \nu_{\rm max}$\fi}
\def\dnobs{\ifmmode \Delta \nu_{\rm obs} \else $\Delta \nu_{\rm obs}$\fi}
\def\ntmax{\ifmmode \nu_{\rm 2max} \else $\nu_{\rm 2max}$\fi}    %
\def\xmax{\ifmmode {\rm X_{\rm max}} \else ${\rm X_{max}}$\fi}

\def\ntobs{\ifmmode \nu_{\rm 2obs} \else $\nu_{\rm 2obs}$\fi}    %
\def\nomax{\ifmmode \nu_{\rm 1max} \else $\nu_{\rm 1max}$\fi}    
\def\nn{\ifmmode \nu_{\rm NBO} \else $\nu_{\rm NBO}$\fi}    
\def\nh{\ifmmode \nu_{\rm HBO} \else $\nu_{\rm HBO}$\fi}    
\def\nqpo{\ifmmode \nu_{QPO} \else $\nu_{QPO}$\fi}    
\def\nz{\ifmmode \nu_{o} \else $\nu_{o}$\fi}    
\def\nht{\ifmmode \nu_{H2} \else $\nu_{H2}$\fi}    
\def\ns{\ifmmode \nu_{s} \else $\nu_{s}$\fi}    %
\def\nb{\ifmmode \nu_{{\rm b}} \else $\nu_{{\rm b}}$\fi}
\def\nkm{\ifmmode \nu_{km} \else $\nu_{km}$\fi}    %
\def\ka{\ifmmode \kappa \else \kappa\fi}    %
\def\dn{\ifmmode \Delta\nu \else \Delta\nu\fi}

\def\vk{\ifmmode v_{\rm k} \else $v_{\rm k}$\fi}    
\def\va{\ifmmode v_{\rm A} \else $v_{\rm A}$\fi}    
\def\vf{\ifmmode v_{\rm ff} \else $v_{\rm ff}$\fi}    

\def\rs{\ifmmode {\rm R_{s}} \else $R_{\rm s}$\fi}    %
\def\ra{\ifmmode R_{\rm A} \else $R_{\rm A}$\fi}    
\def\rso{\ifmmode R_{S1} \else $R_{S1}$\fi}    
\def\rst{\ifmmode R_{S2} \else $R_{S2}$\fi}    
\def\rmm{\ifmmode {\rm R_{M}} \else ${\rm R_{M}}$\fi}    %
\def\ris{\ifmmode {\rm R}_{{\rm ISCO}} \else $ {\rm R}_{{\rm ISCO}} $\fi}
\def\rsix{\ifmmode R_{6} \else $R_{6}$\fi}

\title{The applications  of the MHD Alfv\'en wave oscillation  model for
 kHz quasi-periodic oscillations }

\author{ C.M. Zhang\inst{1}\fnmsep\thanks{Corresponding author:
  \email{zhangcm@bao.ac.cn}}, H.X. Yin\inst{1},
 Y.H. Zhao\inst{1}
\and  H.K. Chang\inst{2} \and L.M. Song\inst{3}}

\institute{National Astronomical Observatories,
 Chinese Academy of Sciences,
Beijing 100012, China \and  Department of Physics and Institute of
Astronomy, National Tsing Hua University, Hsinchu 30013, Taiwan \and
  Astronomical Institute, Institute of High Energy Physics,
 Chinese Academy of Sciences, Beijing 100039, China}


\abstract{
 In this paper,
 we  improve the previous work on  the MHD Alfv\'en wave oscillation model
  for the neutron star (NS)  kHz  quasi-periodic oscillations (QPOs), and
  compare  the model   with the updated twin kHz QPO data.
   For the 17 NS X-ray sources with the simultaneously
detected  twin kHz QPO frequencies, the stellar mass M and radius R
constraints  are given  by means of the derived parameter A in the
model,  which is associated with  the  averaged mass density  of
star as  ${\rm \lan \rho \ran = 3M/(4\pi R^{3}) \simeq 2.4\times
 10^{14} (g/cm^{3}) (A/0.7)^{2}}$, and we also compare the M-R constraints with
 the star  equations  of states.
 Moreover, we also discuss the  theoretical   maximum
  kHz QPO frequency and maximum twin peak separation,
  and  some  expectations  on SAX J1808.4-3658 are mentioned, such as its
 highest kHz QPO frequency  $\sim$~870 (Hz), which is about 1.4-1.5 times less than
 those  of the other known kHz QPO sources.  The  estimated  magnetic fields
 for both Z sources (about Eddington accretion rate  $\med$)
 and Atoll sources ($\sim 1\% \med$)
 are  approximately $\sim 10^9$ G and $\sim 10^8$ G respectively.
}

\maketitle

\section{Introduction}

 The {\it Rossi X-Ray Timing Explorer} (RXTE) has observed
the kilo-Hertz quasi-periodic oscillations (kHz QPOs) in the X-ray
flux in about 25 accreting neutron stars in  low-mass X-ray binaries
 (LMXBs),   and  the  twin peak kHz  QPOs, the  upper and lower
 frequencies ($\nt$ and $\no$), are usually    shown
 in the Fourier power spectrum
 (e.g., van der Klis 2000, 2005, 2006 for the recent reviews).
  Remarkably, the ranges of kHz QPO frequencies  are  almost homogeneous
   at about $\simeq$200 Hz - 1300 Hz  for both   the less luminous Atoll
   sources and the bright  Z sources (on the definition of Atoll and Z, see
  Hasinger \& van der Klis 1989), and they  increase with the inferred mass
  accretion rate,
  which implies that the kHz QPO mechanisms   should be related to
  the common properties of both classes of sources
  (e.g., van der Klis 2000, 2005, 2006; Belloni et al. 2005; Zhang et al. 2006a).
  Furthermore, it is also found
  that the separation of twin kHz QPO peaks  $\dn=\nt-\no$ is not a
  constant and usually decreases with the mass
  accretion rate    (e.g., M\'endez  \& van der Klis 1999;
  van der Klis 2000, 2006).  However, it is found that the peak
  separations increase with the accretion rate when the kHz QPO
  frequencies   are low in Cir X-1 (Boutloukos et al. 2006) and
  in  4U 1728-34 (Migliari  et al. 2003),
  which supports the nonlinear correlations between the
    twin kHz QPO   peaks (e.g. Zhang et al. 2006a),
     and it has been implied by the Alfv\'en oscillation model
     by Zhang  (2004) and  relativistic precession model by  Stella \& Vietri (1999).
Moreover, the kHz QPO  frequencies also follow the rather tight
correlations between themselves and  with the other timing features
of the X-ray emissions (see, e.g., Zhang et al. 2006a; Psaltis et
al. 1998; Psaltis et al. 1999ab; Stella et al. 1999; Belloni et al.
2002; Titarchuk \& Wood 2002),  and the ratios between twin kHz QPO
peaks systematically decrease with the frequency with the averaged
ratio value  of about 1.5 (Zhang et al. 2006a).

  In order to account for  the mechanisms  of twin kHz QPO phenomena,
  some viable models have been proposed.  At the early stage of discovery
  of   kHz QPOs, the  sonic-point beat-frequency model is proposed (see, e.g.,
Miller at al. 1998), which predicts  a constant $\dn$ at the stellar
spin frequency, however, with the further observations, the simple
beat model is inadequate for the varied kHz QPO separations.
 Instead,   in the developed  sonic-point model  by  Lamb \& Miller
(2001),   the authors  consider  the disk flow at the spin-resonant
radius to be  smooth or clumped   to interpret why the occurrence of
kHz QPO pair   separation is close to spin frequency as detected in
XTE J1807-294 (see, e.g. Linares et al. 2005; Zhang et al. 2006b) or
half spin   frequency as detected in SAX J1808.4-3658 (see, e.g.
Wijnands et al. 2003)

Later on, the  relativistic precession model is proposed   by Stella
and Vietri  (1999),  in which the upper  kHz QPO frequency $\nu_2$
is identified with the Keplerian frequency of an orbit in the disk
and the lower  kHz QPO frequency $\nu_1$ with the periastron
precession of that orbit, and the varied $\Delta\nu$ can be
consistently explained in this  model.
 More recently,  Rezania and Samson (2005) propose a
 model to explain the kHz QPOs  based on the interaction of
 accreting plasma with the neutron star magnetosphere, where the
 matter is accelerated by the gravitational pull of the compact
 object and hits the star  magnetosphere with a sonic or supersonic
 speed. Basically the conclusions of this model are consistent with
 the observed data for  the appropriate  choices of the parameters.

Moreover, Titarchuk et al. (1998),  Osherovich \& Titarchuk (1999ab)
and   Titarchuk \& Osherovich (1999) have developed  the alternative
models. They require the lower kHz QPO frequency   to be due to the
Keplerian  frequency  of matter in the disk and the upper kHz QPO
frequency  to be  the hybrid between the lower kHz QPO frequency and
the stellar spin frequency.
 Nonetheless,
 there have not yet been any agreements on the origins of the  QPOs
 in neutron star and black hole binary X-ray sources,
nor on what physical parameters determine their frequencies, which
have also been identified with various characteristic frequencies in
the inner accretion flows (e.g.   Abramowicz et al. 2003ab;
Abramowicz 2005;  Kluzniak et al 2004; Lee et al. 2004; Rebusco
2004;  Rebusco \& Abramowicz 2006;  Petri 2005; Horak \& Karas
2006).

  This   paper is an improved one of the previous work  on the
MHD Alfv\'en wave oscillation mechanism for the NS kHz QPOs  (Zhang
2004), where we consider the radial dependence of the accretion flow
velocity,  and the applications    have been done   by comparing the
model's predictions with  the updated kHz QPO data.

 The paper is  organized as follows: In section 2,  the  overview and
 scenario of the model are   described,  and the derivations of kHz QPO
 frequencies are given. The further  applications of model are presented
 in section 3.
 The  conclusions  and consequences   are summarized  in the final
section.  As the conventional usage, the Newtonian gravitational
constant G and the speed of light c=1 are exploited.

\section{The  overview of the improved  model}

 The previous model  for the NS  kHz QPOs proposed by Zhang (2004)
   does not mention  the excitation mechanism of Alfv\'en wave
   oscillation  to modulate the X-ray flux to produce  the observed  kHz QPOs.
Therefore, in this paper  we try to answer these questions.
  To straightforwardly  grasp  our point of views,
  an imagined geometrical illustration  of
 the  magnetosphere and accretion disk associated with
  our model  is  plotted  in FIGURE~1.

\subsection{MHD wave and its excitation in the accretion flow }

In MHD, the Alfv\'en wave is a transverse wave and propagates along
the  magnetic field line,  the effects of which  have  been studied
in  solar physics to interpret the  detected  quasi-periodic
oscillation   of several minutes in coronal loops (e.g., Roberts
2000;   Nakariakov et al. 1999; Aschwanden et al. 1999).
  As  known,  the coronal loops  may be set into
oscillations with various modes, such as a kink mode (Roberts 2000
for a review), thus we assume  the similar oscillations to
  occur in the accretion disks   with the  loop length of circumference
   $2\pi r$  at  the disk orbit radius r.
As pointed out,  the  MHD turbulence by the   shear flow  in the
accretion disk (e.g.,  Ruediger \& Pipin  2000) will trigger   the
strong variation of plasma energy density and ignite  the   shear
Alfv\'en wave motion along the orbit. In a wave length of  circular
perimeter, the intensely excited  shear Alfv\'en wave can dedicate
to the  observed  X-ray  flux  fluctuations, which is then
responsible  for the observed   kHz QPO phenomena.

 However,  the  propagation of the  Alfv\'en wave will arise its damping
  because of  the dissipation by the X-ray emission
 and the  viscous interactions in MHD,  which may account for why
 sometimes we only measure  a single kHz QPO or nothing.
  From the conventional  accretion disk prescription, the
 accretion flow in the equatorial plane of disk will drag the polar
 field lines  into the azimuthal direction, and it is often
 assumed that the field strength of  the azimuthal component $B_{\phi}$
 is comparable to that of the unperturbed polar field in the equatorial
 plane. 
  i.e. $B_{\phi}(r) \simeq B(r)$
  (e.g., Ghosh \& Lamb 1979; Shapiro \& Teukolsky 1983).

\subsection{On the preferred radius and the coherence}

On the preferred radius r, it is supposed to be a ``coherence"
location at where   the characteristic Alfv\'en velocity $v_{A}(r)$
(frequency) calculated by the plasma mass density of quasi-spherical
flow coincides with the Keplerian velocity $\vk(r)$ (frequency),
i.e., $v_{A}(r)=\vk(r)$.    We assume that at this radius there is a
resonance between the Alfv\'en wave frequency and  Keplerian orbital
frequency.

From the  definition, at the Alfv\'en radius $\ra$,     the
Alfv\'en velocity is the free fall velocity,
 expressed as (e.g., Shapiro \&  Teukolsky 1983),
 \be \label{rhor}
 v_{A}(\ra) = B(\ra)/\sqrt{4\pi\rho(\ra)} = \vf(\ra)\;,
 \ee
 where $B(r)\sim 1/r^3 $ is  the dipole magnetic field
 and
 \be
 \rho(r)=\mdot/[S_{r}\vf(r)]\;,\;\; S_{r}=4\pi r^{2}
\ee
  is the plasma mass
 density   defined by the spherical accretion with the free fall
 velocity \vf(r) or the Keplerian velocity \vk(r),
\be \vf(r) = \sqrt{2GM/r} = \sqrt{2}\vk(r),\;\;
 \vk(r)=\sqrt{GM/r}\;.\ee
   Therefore, at the preferred radius  $r = \phi \ra$, where
$\phi$ is an introduced parameter,
 the Alfv\'en velocity   equals the Keplerian velocity, thus,
\be v_{A}(r) = B(r)/\sqrt{4\pi\rho(r)}
 = \sqrt{2}\phi^{-7/4}\vk(r) = \vk(r)\,, \label{va}\ee
  the   `coherence'
radius is defined by the condition of \\$\sqrt{2}\phi^{-7/4} = 1$
 or $\phi=2^{2/7}\simeq1.2$.

\subsection{The identifications of twin kHz QPO frequencies}
Therefore, at the preferred radius,  the Alfv\'en wave frequency
is \be \nu_{A}(r) = {v_{A}(r)\ov 2\pi r} = {\vk(r)\ov 2\pi r} =
\nk(r)\;,
 \ee which we identify to the
  upper kHz QPO frequency (see also Zhang 2004),  or  written as
 \bea \label{nt}
  \nt = \nk &=& 1850 {}(\xi A){}X^{3/2} ({\rm Hz}) \\ \nonumber
 &=& 1295 ({\xi A\ov 0.7}){}X^{3/2} ({\rm Hz})\,,  \eea
 with  the parameters  X=R/r and
 $A = (m/R_{6}^{3})^{1/2}$  where   $R_{6}=R/10^{6}(cm)$
 and   $m={\rm M \over \ms}$  are the stellar  radius R and  mass M
  in  the units  of $10^{6}$ (cm) and solar masses, respectively.
  It is noted that the quantity  $A^{2}$ is a measurement of the averaged
  mass density of star, expressed as \\
  ${\rm \lan \rho \ran = 3M/(4\pi R^{3}) \simeq 2.4\times 10^{14} (g/cm^{3})
  (A/0.7)^{2}}$. \\
  The reasons for introducing the ``non-Keplerian factor" $\xi$ in
 Eq.(\ref{nt}) come  from the facts of the complicated physical environments,
  which do not satisfy  the ideal conditions for applying the Keplerian frequency,
  i.e. the point mass orbiting around a central gravitational source
  in a vacuum. The modifications to the Keplerian frequency  will be
  taken into account if the following situations are considered:
   (a) the  influence by rotation or the modification from Kerr spacetime
  to Schwarzschild spacetime;
  (b) the consideration of the factual environment around star deviating
  from the ideal situation of a test particle motion in a vacuum;
   (c) the plasma blob moving in a MHD with the strong magnetic
   field.
  In principle, this ``non-Keplerian factor" $\xi$ should be not
  a multiplicative constant. For instance, it could depend upon the
  radius in the disk at which the QPO is produced, and hence upon
 the frequency itself. However, we have to stress that it is just
 the reason of simplicity to  chose it as a constant.
     In practice,  for the mathematical convenience, we still take $\xi=1$ to
   process  the calculations, and finally
   the real parameter A  can be obtained  through dividing it by
    the  ``non-Keplerian factor"   $\xi$, the implication of
   which is discussed in the last section.

 Equivalently, the NS radius can be  expressed by
the parameters A and m, \be
 R_{6} = 1.27{}m^{1/3}(A/0.7)^{-2/3} {\,} (10~ {\rm km})\; .
\label{mr} \ee
 Furthermore, on the formation scenario of the lower kHz QPO
  frequency ($\no$), we have the following arguments.
  The  accreted materials accumulated on the polar cap will become
  denser by a factor of ratio of the spherical area to the polar cap area.
 These denser materials will flow equator-ward and pass through the
 perpendicular field lines by the instabilities, so this
 flow is influenced by the Lorentz  force.
   The all denser materials will be expelled out   through the
 magnetic tunnel  and
 enter   into   the  transitional zone  near the `coherence' radius
 because the flow along the field lines is force free and will not
 experience the Lorentz force.
  Henceforth, we ascribe the Alfv\'en wave frequency calculated
   by  the denser mass density of plasma formed  on the polar cap
    to be  the lower kHz QPO frequency.
  For the mathematical convenience,
 it is presumed that  both  the upper and lower kHz QPOs
 occur  at the same radius.
 Thus, the lower kHz QPO frequency $\no$ is described below,
 \be \label{nuap}
 \nu_{Ap}(r) =  {v_{Ap}(r) \over 2\pi r },  \ee
 where the Alfv\'en
velocity   $v_{Ap}(r)$ is defined by the mass density expelled
from the polar cap, \be v_{Ap}(r) = \frac{B(r)}{\sqrt{4\pi
\rho_{p}}} = \frac{B(r)}{\sqrt{4\pi
\rho(r)}}\sqrt{{\rho(r)\ov\rho_{p}}} =
v_{k}(r)\sqrt{{\rho(r)\ov\rho_{p}}}\;, \ee with
\be\rho_{p}=\mdot/[S_{p}\vf(R)]\,,\,\,\, 
\ee where  the magnetic polar cap area $S_{p}$ is  obtained  as
(Zhang \& Kojima 2006),
 \be S_{p}=4\pi R^{2} (1 - cos\theta)\, ,\label{sp}\ee
 with
 \be \sin^{2}\theta={R/r}\equiv X \, , \ee where $\theta$ is
the open angle of the last field line to close at radius r. As an
approximation, the polar cap area is usually  written as
$S_{p}=\frac{2\pi R^{3}}{r}$ if $R \ll r$
 (see e.g. Shapiro \& Teukolsky 1983, P.453).
  Henceforth,
 if we ascribe the upper and lower  kHz QPO frequencies to
  the two Alfv\'en wave frequencies with different mass
  densities  described in Eqs.(\ref{nt}) and (\ref{nuap}),
   then we have

\be {\no\ov\nt} =
\sqrt{{\rho(r)\ov\rho_{p}}}=\sqrt{{S_{p}\vf(R)\ov S_{r}\vf(r)}}\;.
\label{nu1to2}\ee After considering Eq.(\ref{rhor}) and
Eq.(\ref{sp}) with the parameter definition X$\equiv$ R/r,
Eq.(\ref{nu1to2}) gives
  \be \no
 = \nt X^{-1/4}\sqrt{{S_{p}\ov S_{r}}} = \nt
 X^{3/4}[1-\sqrt{1-X}]^{1/2}\;. \label{no}
\ee
For convenience, the ratio of  the twin kHz QPO frequencies  is
obtained to be,
\be
{\nt \over \no} = X^{-5/4} \sqrt{1 + \sqrt{1-X}}\,, \label{ratio}
\ee   which only depends on the position parameter X$\equiv$ R/r
where X-ray flux responsible for  kHz QPOs  emits  and is
independent of the averaged mass density parameter A  and mass M.
Furthermore, the twin kHz QPO separation is written as, \be \dn
\equiv  \nt - \no =
 \nt[1 - X^{3/4}(1 - \sqrt{1-X})^{1/2}] \,,
\label{separation} \ee
which is not a constant with the variation
of $\no$ or $\nt$.

From Eq.(\ref{nt}) and Eq.(\ref{no}), if the twin kHz QPO
frequencies are known simultaneously, then the values of  A and X
can be determined. For the  detected  sample  sources listed in
TABLE I (see e.g. van der Klis 2000, 2006; Belloni et al. 2005; data
are provided by T. Belloni, M. M\'endez and D. Psaltis), such as Sco
X-1 (van der Klis et al. 1997; M\'endez \& van der Klis 2000),
4U1608$-$52 (M\'endez et al. 1998ab) , 4U1735-44 (Ford et al. 1998)
and 4U1728$-$34 (M\'endez \& van der Klis 1999).

 The comparisons of the model's conclusions to the
well  detected kHz QPO sample  sources  are shown in FIGURE 2, and
the agreement  between the model and the observed QPO data is quite
good for the   selected ranges of  NS parameters A=0.6, 0.7 and 0.8.
In FIGURE 2b, we find that $\dn$ increases with $\nt$ if $\nt <
\sim750$ (A/0.7) Hz and $\dn$ decreases with $\nt$ if $\nt >
\sim750$ (A/0.7) Hz. In addition, the twin kHz QPO ratio (FIGURE 2c)
follows the decreasing tendency with $\nt$.

\section{The  applications of the model}

To inspect  the model's predictions, we
 demonstrate some applications  of the model in the following.

\subsection{The constrain conditions of NS  mass and radius}

The NS mass  constrain condition  by the kHz QPOs
 has  been  given by Miller et al (1998) and  Zhang et al (1997)
 through  assuming that the  accretion disk  radius    of showing  kHz
 QPOs
 is bigger   than the innermost stable circular orbit (ISCO:
 three Schwarzschild radii),
  i.e., the maximum observed frequency is presumed to be the   Keplerian
frequency at ISCO  with the condition that the stellar  surface is
enclosed by ISCO, \be m \leq   2.2/\nu_{2k}\,. \label{millereqm} \ee
From   Eq.(\ref{mr}), the new radius constrain condition is obtained
if the mass constrain condition is known, therefore we exploit the
generally assumed  NS mass lower limit
 1.0 $\ms$  from the astrophysical argument   and its   upper
limit condition Eq.(\ref{millereqm}) to  set the radius  constrain
conditions, respectively,
\be R_{6} \geq 1.27(A/0.7)^{-2/3}\,,
\label{rrangel}%
\ee
and
\be
R_{6} \leq  ({2.2 \over \ntk A^{2}})^{1/3} =
 1.65 \nu_{2k}^{-1/3}({A\over0.7})^{-2/3}\,.
\label{rrange}%
\ee
 In Eq.(\ref{rrange}), the radius is constrained by A and
$\nt$, or equivalently by the twin kHz QPOs.
  Furthermore,  the NS mass and radius relations and their constrain conditions
 are calculated  for the 17 known  sources whos
twin kHz QPOs are   simultaneously detected, which are  listed in
TABLE I. In Figure~3, the mass-radius relations inferred from the
values of parameter A  have been plotted, and it is found that, for
the conventionally accepted NS mass lower limit $\sim\ms$, there
exists difficulty in accordance with the many modern equations of
states (EOSs) if the parameter A is too low, for instance
$A\sim0.47$  for SAX J1808.4-3658 (see TABLE I). In general, the
lower the value of A, the more difficult it is to reconcile the M-R
relations with the realistic EOSs of stars. The situation can be
altered if the introduced  `non-Keplerian factor'  $\xi<$ 1, which
makes the parameter A increased.

\subsection{The maximum twin kHz QPO frequencies and their
separations} The theoretical maximum frequency separation $\dnmax$
can be calculated  from Eq.(\ref{separation}) by vanishing the
variation of $\dn$ respect to X, $d[\dnmax]/dX = 0$, where we
obtain $X\simeq0.69$, $\nt = 750 (A/0.7) ({\rm Hz})$, $\no = 382
(A/0.7) ({\rm Hz})$ and $\dnmax = 368{} (A/0.7) ({\rm Hz})$. If
$\nt$ is lower (higher) than ~750 (A/0.7) $({\rm Hz})$,
 then $\dn$ will increase (decrease) with $\nt$.
 In the recent observations of 4U 1728-34 (Migliari et al 2003),
the  twin  simultaneously detected  kHz QPO frequencies are found
 at the central frequency $\no = 308$ (Hz) and
 $\nt = 582 $ (Hz) ( $\dn = 274 $ (Hz)),
 so this is the first detected event for a significant
 decrease of kHz QPO peak separation towards low frequencies,
 however,
which is qualitatively consistent with our model's prediction.
Nonetheless, the maximum values of twin   kHz QPO frequencies
coincide  and  occur at X=1 where the preferred radius equals
the stellar  radius in case the ISCO is inside the star,
 $\nomax=\ntmax=1850{}A ({\rm Hz})$ =
$1295(A/0.7) ({\rm Hz})$, namely \be A \ge {\rm \nk\ov 1850
(Hz)}\;. \label{a} \ee
 In  TABLE I, the theoretical maximum kHz QPO frequencies of 17
  kHz QPO sources  are   listed, and
the averaged  values of A  and the maximum position parameter
$\xmax$ are given for both the Z and Atoll sources, which are,
respectively,
 $\lan A \ran$=0.66 and  $\lan \xmax \ran$=0.87  ($r\sim1.15 R$)
 for Z sources,  and
$\lan A \ran$=0.74 and  $\xmax$=0.9  ($r\sim1.1 R$) for Atoll
sources.  However,
  we currently  cannot figure ~ out  what physical mechanisms
  arise  these  systematical differences between the Z and Atoll sources.
 The  inequality
   $\xmax < 1$ tells us the fact that the accretion disk does not reach the
 stellar surface,
 so it is likely that the ISCO prohibits the disk from
 arriving at the stellar surface. 
If this scenario is plausible, we can conclude that the ISCOs of
  almost all  sources in TABLE I  locate outside  the stars and the
 estimated positions of their masses and radii
 should appear above ISCO line in  M-R diagram shown in FIGURE~3.
 Interestingly, the  one unusual case  is the Atoll source 4U 0614+09
 (e.g., van Straaten et al 2000; van der Klis 2000, 2006),
 and its parameter $A \simeq  0.76$ is
inferred  from the simultaneously detected twin kHz QPOs, which
 implies its maximum  upper (lower) kHz QPO frequency  to be  about 1406 (Hz).
 However, from the observation, one single kHz QPO peak frequency 1330 (Hz)
 was detected in  this source (van Straaten et al 2000;
van der Klis 2000, 2006), showing $ X \sim 0.97$,  thus this
detected single frequency  may  be near the maximum  kHz QPO
saturation frequency,
   showing   the X-ray spectrum information near  the stellar  surface,
  the confirmation of which needs the further proposal  and analysis
  of this source.

  As for the level-off or  frequency saturation  of kHz QPO,
 it  has been
paid much attention since the early discovery of kHz QPOs,  which is
ascribed to the occurrence of ISCO or stellar surface  (e.g., Zhang
et al 1998; Kaaret et al 1999; Miller 2004; Swank 2004). However,
from our model,  if $\ris > R$, the saturation frequency would
 occur at ISCO with the maximum upper kHz QPO frequency
 $\ntmax = 2200/m$ (Hz) (see also, e.g. Miller et al  1998; Miller 2004);
 if $\ris < R$, the saturation frequency
 would occur at the stellar  surface R with  $\ntmax = 1295(A/0.7)$ (Hz).
  Therefore, the model inferred maximum positions  of    showing   the kHz
  QPOs    satisfy
 $\xmax < 1$ (see TABLE I) for the  simultaneously detected kHz QPO
sources, which   implies that either their   ISCOs   appear outside
stars or the unclear  mechanisms diminish  the kHz QPO X-ray spectra
just above the stellar  surface,  which needs the further
investigations.

\subsection{The estimations of magnetic field strengths
 of NSs in Z and Atoll sources}

The estimation of the NS magnetic field strength $B$
 can be  given by the definition of its magnetosphere (see, e.g.
 Shapiro \& Teukolsky 1983), or described by
 the  accretion induced magnetic evolution model
 (see, e.g. Zhang \& Kojima 2006; Cheng \& Zhang 1998),
 \be B = ({\rmm\over R})^{7/4}\bff\;,
 \label{bfield}
 \ee
 \be\bff \simeq 4.3\times10^{8}\; (G)\;
 (\la\mdot\rangle/\med)^{1/2}m^{1/4}R_{6}^{-5/4}\;, \ee where $\rmm$
is the magnetosphere radius defined by the long-term accretion rate
$\la\mdot\rangle$ and  $\med$ is Eddington limited accretion rate.
kHz QPOs are assumed to  be produced around $\rmm$ with the
variation of the instantaneous accretion rate, thus  $R/\rmm$ should
be   comparable to the averaged position parameter $\la X\rangle$,
i.e. $\la X\rangle\sim R/\rmm$,  which can be estimated  by
Eq.(\ref{nt}) as \be \la\nt\rangle \simeq 1850A\la X\rangle^{3/2}\;,
\ee where $\la\nt\rangle $ is the averaged value of the detected
upper kHz QPO of Z ($\sim$870 Hz) or Atoll ($\sim$980 Hz) sources
 and both sources share the  homogeneous kHz QPO distributions (see,
e.g. Zhang et al. 2006a), therefore, \bea B
&\simeq& (1850A/\la\nt\rangle)^{7/6}\bff\\
& \simeq & 10^{9} (G)\; ({\la\nt\rangle\ov900 {\rm Hz}})^{-7/6}
 ({\la\mdot\rangle\ov \med})^{1/2}m^{5/6}R_{6}^{-3}\;.
 \eea
 So, the   magnetic fields are proportionally related to the
 accretion rates, and for both Z sources (Eddington accretion rate
 $\med$) and Atoll sources ($\sim 1\% \med$) they
 are about $\sim 10^9$ G and $\sim 10^8$ G respectively
 if $ m\sim 1$ and $R_{6}\sim 1$,
which are  consistent with the originally hinted  values from the
X-ray spectra of both sources (Hasinger \& van der Klis 1989).

On  the correlation between QPO frequency and the accretion rate, we
can write it in the following by means of the definition of
magnetosphere (Shapiro \&  Teukolsky 1983), \bea
 \nu_{2} &   \sim &  (R/r)^{3/2} \sim (R/\rmm)^{3/2}(\rmm/r)^{3/2} \\
 &  \sim  & ({\la\mdot\rangle\ov B^{2}})^{3/7}
 ({\mdot\ov \la\mdot\rangle})^{3/7}
  \sim   ({\la\mdot\rangle \ov B^{2}})^{3/7} \mxd^{3/7}\;,\eea where
$\mxd=\mdot/\la\mdot\rangle$ is a ratio of the instantaneous
accretion rate to the long-term accretion rate. If the neutron star
accreted $\sim 0.01\ms$, the NS magnetic field will enter into a
`bottom state' where the B-field is proportionally related to the
accretion rate as $B\propto\la\mdot\rangle^{1/2}$ by the accretion
induced magnetic evolution model (Zhang \& Kojima 2006). Thus, we
obtain a unified expression of QPO frequency vs. the accretion rate
for both Atoll and Z sources to be $\nt\sim \mxd^{3/7}$, although
both sources share a diversified luminosity of even more than two
magnitude orders.

\section{Summaries  and conclusions}

 In the paper, we  compare the observations with
  the improved model for kHz QPOs, and the   main conclusions are
summarized in the following.

 (1) The theoretical
 maximum  twin kHz QPO frequencies coincide and occur at
   about 1295(A/0.7) (Hz) when the accreted matters to    show
 these QPOs clash  on the NS  surface, and
 the maximum  twin kHz QPO  separation is 368 (A/0.7) (Hz).
(2) For SAX~J1808.4-3658,
 we find that   its stellar mass density parameter A=0.47 is
  about  1.5 times less than the typical  values of
  other kHz QPO  sources with  A$\sim$0.7,
 and  we also obtain its highest kHz QPO frequency
to be  $\sim$870 (Hz), which needs the proof of future detections.
 (3)  The averaged mass density of NS
can be described  by the defined   parameter A as\\
 ${\rm \lan \rho \ran = 3M/(4\pi R^{3})
  \simeq 2.4\times10^{14} (g/cm^{3}) (A/0.7)^{2}}$,
  and we obtain ${\rm \lan \rho \ran \simeq 2.4\times
 10^{14} (g/cm^{3})}$  for most of NS kHz QPO sources with
 $A\simeq0.7$.
(4) With  the derived  parameter A  listed in
   TABLE I for the ideal case of
   the    ``non-Keplerian factor"   $\xi=1$, the mass-radius
   relation curves are  plotted in FIGURE~3,
 and we find that the  EOSs  of strange matters
(CS1 and CS2) seems to be not favorite except $A \sim  1$, resulting
in an extremely high kHz QPO frequency $\ntmax \sim 1800$ (Hz).
 Moreover, for the generally assumed
 NS mass lower limit, $\sim \ms$ for instance, EOSs of
 the normal neutron matters (CN1 and CN2) do not fit for
  the stars in the detected kHz QPO sources
  unless  $A \sim 0.88$, corresponding to
 $\ntmax \sim 1600$ (Hz). If  the EOSs of CPC
(the star core becomes a Bose-Einstein condensate of
 pions) are   the possible choices,   the mass and radius ranges
  of these stars   are  from  $1.0 \ms$   to $1.7 \ms$  and from
   15 km to 18 km, respectively,
for A = 0.45 -- 0.79.  In addition, if the introduced
``non-Keplerian factor"   $\xi$ is less than unity, for instance
$\xi\simeq0.7$, then the derived  values of A in TABLE I will be
increased by a factor of about 1.4, for instance from $A\simeq0.7$
to $A\simeq 1.0$, which in turn, as shown in FIGURE ~3, results in a
fact that many sources listed in TABLE I may be the candidates of
strange stars (see, e.g., Cheng et al 1998; Lattimer \& Prakash
2004). However, in our model it seems to be difficult to take the
star in
 SAX J1808.4-3658  (its A=0.47) as   a candidate of strange star as
expected by Li et al. (1999), as seen in FIGURE~3. Nevertheless, the
EOSs of stars in kHz QPO sources are still the open  issues before
the physical influences of the  ``non-Keplerian factor"   $\xi$ are
thoroughly settled.
 (5) From the homogeneous  kHz QPO distributions for both  Atoll and Z
 sources (see, e.g. Zhang et al. 2006a),
 we conclude the NS magnetosphere scales of both sources to be similar,
 which will arise the NS magnetic field to be proportionally
 related to the long-term averaged accretion rate, i.e. NS in Z
 source possesses a stronger magnetic field than that of Atoll
 source.

Furthermore, the parameter A = 0.78  is implied by the model for
KS\,1731-260, corresponding to the maximum kHz QPO frequency
  1443 (Hz), which is bigger than the
 known detected  maximum value  $\nt \sim 1330$~(Hz)
(4U~0614+091,  van Straaten et al. 2000). Moreover, the proposal for
the detection of 1500 (Hz) and 1800 (Hz) QPO frequency is suggested
 by Miller (2004),  corresponding  to  $A \sim 0.8$  and  $A \sim
 1.0$ in our model,  respectively,  so these measurements of QPOs
above 1500~(Hz) therefore have excellent prospects for stronger
constraints on the mass and radius relations, as well as on the
models.  It is  claimed by Miller (2004)  that  a  QPO frequency as
high as 1800~(Hz) would be large enough to argue against all
standard nucleonic or hybrid quark matter EOS, leaving only strange
stars (see also FIGURE~3).  Nevertheless,  the proposal of detecting
the    QPO data at low frequency $\no \sim$ 100 (Hz) is also
meaningful, by which the model is tested through  inspecting $\dn$
vs. $\nt$ relation as shown in the middle panel of  FIGURE 2. It is
also predicted by Stella and Vietri (1999) $\dn$ will increase with
the accretion rate if $\nt < \sim 700$ Hz,
 and the evidence for this has been recently detected in Cir X-1 by
 Boutloukos et al. (2006).

As a summary, it is remarked that our  model is still a simple one,
and the further improvements are still needed  through considering
the details of accretion flow or disk structure,  where the
plausible existence  of the transitional layer (e.g., Titarchuk et
al 1998; Titarchuk \& Osherovich 2000)  or the  nonlinear disk
resonances (e.g., Abramowicz et al. 2003ab) may be important.
 In addition,    we  have not yet considered
 the rotational effect of  Kerr spacetime (see, e.g.,
  Miller 2004; van der Klis 2006), the  factual  MHD flow velocity
 around star but perhaps  not a Keplerian velocity of test particle,
 the accretion flow   in a strong gravity regime,
 the plasma instabilities and turbulence  in strong magnetic field, etc.
 Therefore,   the considerations of above complications
 will improve the present version of the model.

\vskip .3cm

\acknowledgements

We thank  T. Belloni, M. M\'endez,  D. Psaltis  and M.C. Miller for
providing  the data files,  and  helpful discussions with
 T.P. Li, X.D. Li, J.L. Qu, S.N. Zhang, M. Abramowicz, S. Boutloukos, J.
 Horak, J. Homan, V. Karas,  P. Rebusco and J. Petri  are highly appreciated.
 This  research has been supported by the innovative  project of CAS of China.
 C.M. Z. thanks MPE-Garching and  TIARA-NTHU for visiting supports.
 H.-K. C. is supported by the National Science Council through grants
NSC 94-2112-M-007-002 and NSC 94-2752-M-007-002-PAE.
%
 We are very grateful for the critical comments and helpful
 suggestions  from the anonymous referee, which advise us to thoroughly
  improve  the quality of the paper.

\newpage

\begin{figure}\label{nss}
\begin{center}
\includegraphics[width=7.5cm]{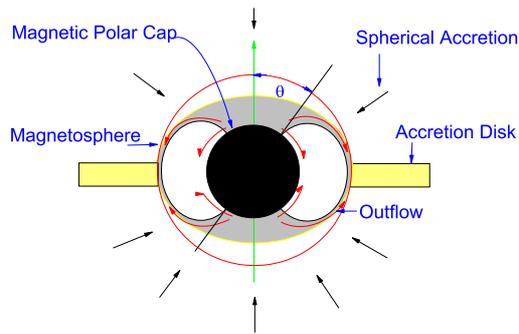}
\vskip 2.5cm \hskip   -6.80cm 
\end{center}
\caption{ An imagined  schematic illustration of an accreting
neutron star magnetosphere and disk for the kHz QPO productions.
$\theta$ is the open angle between the magnetic axis and the closed
field line of magnetosphere, defined by $\sin^{2}\theta=R/r$, where
R and r are the radii of star and its magnetosphere, respectively.
 The quasi-spherical accretion from all over star falls in and then
cumulates at
 the magnetic polar cap, where the blobs or patches of
accreting plasmas are piled up and condensed into the high mass
 density, and some of
 materials
will   diffuse   onto the whole stellar area  through the side-flow
on account of  the plasma instabilities and the others sprout out
along the closed field lines through the
  out-flow  to enter into the orbital  flow.
  The preferred radius is supposed at where
the shear Alf\'en
 wave frequency with the quasi-spherical accretion
 propagating in  the orbit equals
  the orbital Keplerian frequency, namely the
 ``coherence" is presumed to occur there.
    }
\end{figure}

\begin{figure}\label{nu12}
\includegraphics[width=7.5cm]{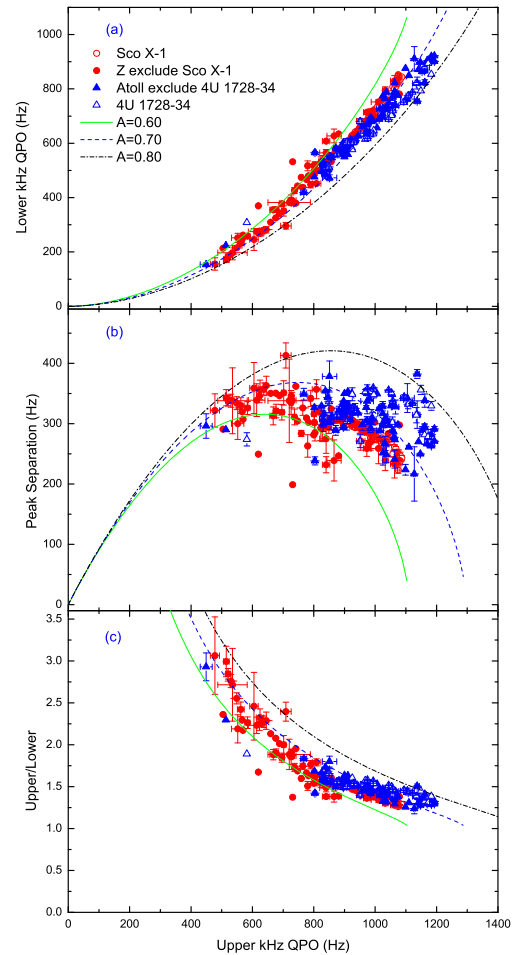}
\vskip 2.5cm \hskip   -.42 cm 
\caption{  Plots of (a) $\no$ vs. $\nt$,  (b) $\dn$ vs. $\nt$ and
(c) $\nt/\no$ vs. $\nt$. The choices of the parameters and the kHz
QPO samples are indicated in the figure. The data are provided  by
T. Belloni, M. M\'endez and D. Psaltis. }
\end{figure}

\begin{figure}\label{fmr}
\includegraphics[width=7.5cm]{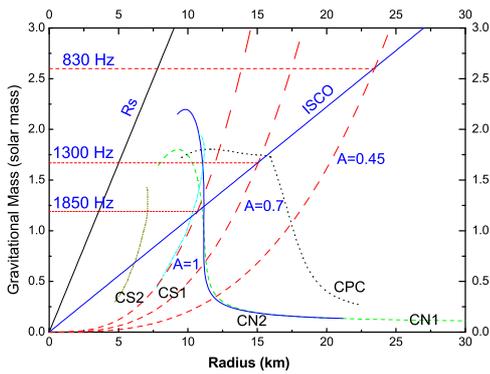}
\vskip 2.5cm \hskip   -.40cm 
 \caption{
The mass versus radius diagram. The straight lines labeled by $\rs$
(ISCO) represents the radius equal to one (three) Schwarzschild
radius, R=$\rs$=2GM ($\ris$=6GM), below which R$>$2GM (6GM) is
satisfied.
 The three  parabola curves, from left to right,  represent
the  different parameter conditions with A = 1, 0.7 and 0.45,
respectively, and the corresponding maximum kHz QPO frequencies are
indicated with the  horizontal lines (see also, e.g., Miller et al.
1998). Above (below) the horizontal lines the stellar surface is
inside (outside) the ISCO, so in this case the maximum kHz QPO
frequency occurs at r=$\ris$ (r=R). The five groups of mass-radius
relations for  various NS  EOSs are shown (the data were kindly
provided by M.C. Miller, the original references therein), which is
after M.C. Miller (2002): stars containing strange matter (CS1 and
CS2); stars made of normal neutron matter (CN1 and CN2); and stars
whose cores become a Bose-Einstein condensate of pions (CPC). }
\end{figure}

\newpage

\clearpage
\newpage

\vskip 1.cm
\begin{tabular}{c}
{TABLE I. Neutron star  parameters of  detected kHz QPO sources \hfill}\\
\end{tabular}
\vskip 0.1cm
\begin{tabular}{cccccccc}
\hline\hline
Source$^{*}$      & A$^{(1)}$    & R$^{(2)}$    & m$^{(3)}$
& R$^{(4)}$  & $\ntmax$$^{(5)}$ & $\ntobs$$^{(6)}$ &    $\xmax^{(7)}$ \\
 & & $m^{1/3}$ (km) & ($\ms$) & (km)  & (Hz) & (Hz)&  \\\hline
\hline
{\bf Millisecond pulsar}  \\
\hline
SAXJ\,1808.4-3658$^{a}$  & 0.47    & 16.9 &  1.0-2.99  & 16.8-24.4  & 870  &725 & 0.90   \\
\hline
{\bf Z source}  \\
\hline
Sco~X-1   & 0.66   & 13.3   & 1.0-2.05   & 13.3-17.0 & 1221 & 1075  & 0.93 \\
GX\,340+0 & 0.64   & 13.5   & 1.0-2.62   & 13.5-19.0 & 1184 & 840   & 0.80 \\
GX\,349+2 & 0.68   & 13.0   & 1.0-2.23   & 13.1-17.2 & 1258 & 985   & 0.86 \\
GX\,5--1  & 0.63   & 13.8   & 1.0-2.49   & 13.8-18.8 & 1166 & 890   & 0.84 \\
GX\,17+2  & 0.67   & 13.2   & 1.0-2.02   & 13.3-16.8 & 1240 & 1087  & 0.93 \\
Cyg\,X-2  & 0.70   & 12.8   & 1.0-2.19   & 12.8-16.7 & 1295 & 1005  & 0.85 \\
\hline
{\bf Atoll  source}  \\
\hline
4U\,0614+09  & 0.76   & 12.2   &  1.0-1.65  & 12.1-14.4  & 1406 & 1330 & 0.97 \\
4U\,1636--53 & 0.74   & 12.4   &  1.0-1.79  & 12.1-15.1  & 1369 & 1230 & 0.94 \\
4U\,1608$-$52& 0.69   & 12.9   &  1.0-2.00  & 12.8-16.2  & 1277 & 1099 & 0.90 \\
4U\,1702--43 & 0.75   & 12.3   &  1.0-2.03  & 12.2-15.6  & 1388 & 1085 & 0.85\\
4U\,1728$-$34& 0.77   & 12.0   &  1.0-1.88  & 11.9-14.8  & 1425 & 1173 & 0.88 \\
KS\,1731--260& 0.78   & 11.9   &  1.0-1.83  & 11.9-14.6  & 1443 & 1205 & 0.89\\
4U\,1735$-$44& 0.73   & 12.5   &  1.0-1.91  & 12.4-15.5  & 1351 & 1150 & 0.90 \\
4U\,1820--30 & 0.73   & 12.5   &  1.0-2.00  & 12.4-15.8  & 1351 & 1100 & 0.88\\
4U 1916--053 & 0.73   & 12.5   &  1.0-2.08  & 12.4-16.0  & 1351 & 1058 & 0.85\\
XTE\,J2123--058& 0.72 & 12.6   &  1.0-1.93  & 12.5-15.8  & 1332 & 1140 & 0.91\\
\hline
\hline
\end{tabular}\\

\vskip 0.3cm

{
\begin{tabular}{c}
\begin{minipage}{1.85\linewidth}
\noindent *: The sources are chosen from those that
  the twin kHz QPOs are detected simultaneously
(van der Klis 2000, 2006;  the  original references therein),
 then 4U\,1915--05 is not included because its
  two incompatible values of $\dn$ are reported (van der
Klis 2006).
a: Wijnands et al. 2003. $^{(1)}$: Calculated by the simultaneously
detected twin kHz QPO data. $^{(2)}$: Obtained by Eq.(\ref{mr}).
$^{(3)}$: Estimated by the  generally assumed   NS mass lower limit
1.0 $\ms$ and by the constrain condition $m\leq2.2/\ntk$ (Miller et
al. 1998). $^{(4)}$: Obtained by Eq.(\ref{rrangel}) and
Eq.(\ref{rrange}).
 $^{(5)}$: $\ntmax=1295 (A/0.7)$ (Hz) is the possible maximum kHz QPO frequency.
 $^{(6)}$: $\ntobs$ is the  detected maximum kHz QPO frequency.
 $^{(7)}$: the detection inferred   maximum X position,  $\xmax=(\ntobs/\ntmax)^{2/3}$.
\end{minipage}
\end{tabular}
}

\end{document}